\documentclass[12pt,onecolumn]{IEEEtran}

\usepackage{lineno}
\usepackage{graphicx,color}
\usepackage{amssymb,amsmath}
\usepackage{MnSymbol}
\usepackage{algorithm}
\usepackage{algorithmic}
\usepackage{amsthm}
\usepackage[numbers]{natbib}
\usepackage{setspace}
\modulolinenumbers[5]

\newtheorem{theorem}{Theorem}
\newtheorem{lemma}{Lemma}
\newtheorem{proposition}{Proposition}
\newtheorem{corollary}{Corollary}
\newtheorem{definition}{Definition}
\newtheorem{example}{Example}
\newtheorem{remark}{Remark}
\newcommand{\F}{\mathbb{F}}

\newcommand{\Fqm}{\mathbb{F}_{q^m}}
\newcommand{\Fq}{\mathbb{F}_q}
\newcommand{\xs}{[x;\sigma]}

\newcommand{\cI}{\mathcal{I}}

\newcommand{\cA}{\mathcal{A}}
\newcommand{\cF}{\mathcal{F}}
\newcommand{\cP}{\mathcal{P}}
\newcommand{\dcF}{d_\cF}
\newcommand{\vp}{\varphi}
\newcommand{\vps}{\varphi_{\sigma_s}}
\DeclareMathOperator{\grcd}{grcd}
\DeclareMathOperator{\llcm}{llcm}

\title{Matroidal Structure of Skew Polynomial Rings with Application
  to Network Coding}

\author{
\IEEEauthorblockN{Siyu~Liu$^*$, Felice~Manganiello$^\S$, and Frank~R.~Kschischang$^*$ \\}
\IEEEauthorblockA{$^*$Department of Electrical and Computer Engineering, University of Toronto, Canada \\
$^\S$Department of Mathematical Sciences, Clemson University, United States \\}
Emails: $^*$\{siyu, frank\}@ece.utoronto.ca, $^\S$manganm@clemson.edu
}

\begin{document}

\maketitle

\begin{abstract}
Over a finite field $\Fqm$, the evaluation of skew polynomials is
intimately related to the evaluation of linearized polynomials. This
connection allows one to relate the concept of polynomial independence
defined for skew polynomials to the familiar concept of linear
independence for vector spaces. This relation allows for the
definition of a representable matroid called the
$\Fqm[x;\sigma]$-matroid, with rank function that makes it a metric
space. Specific submatroids of this matroid are individually
bijectively isometric to the projective geometry of $\Fqm$ equipped
with the subspace metric. This isometry allows one to use the
$\Fqm[x;\sigma]$-matroid in a matroidal network coding application. 
\end{abstract}

\section{Introduction}
In numerous recent works, skew polynomial rings have been used to construct algebraic codes \cite{Ulmer_1,Ulmer_2, Ulmer_3,Heide}, for decoding algorithms \cite{Sidorenko, Liu}, and for cryptographic applications \cite{Boucher, Zhang}. 

Early works in \cite{Ore, Cohn, Jacobson} examined the algebraic properties of skew polynomial rings. In the seminal work of Lam and Leroy \cite{Lam2}, a natural way to define an evaluation map on skew polynomial rings was introduced. In addition, associated to this evaluation map, the notion of $\sigma_s$-conjugacy classes, minimal polynomials, and polynomial independence ($P$-independence) were also introduced. 

In this work, we consider the evaluation of skew polynomials defined over a finite field $\Fqm$. In this special case, skew polynomial evaluation is deeply connected to the evaluation of linearized polynomials over $\Fqm$ \cite{Lidl}. It is well known that the evaluation of a linearized polynomial is a linear map. Using this, we give a simple proof of a structure theorem relating the concepts of $P$-independence and linear independence when restricted to a single $\sigma_s$-conjugacy class. 

This structure theorem allows us to define a representable matroid called the $\Fqm[x;\sigma]$-matroid. Using a decomposition theorem on minimal polynomials, we show that the rank function on the $\Fqm[x;\sigma]$-matroid is in fact a metric, thereby making the $\Fqm[x;\sigma]$-matroid a metric space. In particular, specific submatroids of the $\Fqm[x;\sigma]$-matroid are individually bijectively isometric to the projective geometry of $\Fqm$ equipped with the subspace metric defined in \cite{Kschischang}. This isometry allows us to use the $\Fqm[x;\sigma]$-matroid in the matroidal network coding framework defined in \cite{Gadouleau}. 

The rest of this paper is organized as follows. Section \ref{s:skew} discusses some basic properties of skew polynomial rings, with emphasis on defining an evaluation map, the notions of $\sigma_s$-conjugacy classes, and the connections to linearized polynomials. Section \ref{s:indep} introduces the concepts of minimal polynomials and $P$-independence and  states and proves the main structure theorem. Section \ref{s:matroid} introduces matroids and shows that the structure theorem from Section \ref{s:indep} gives rise to a representable matroid. Section \ref{s:comm} describes the application to matroidal network coding and discusses some computational complexity issues involved in this communication model. Section \ref{s:conclusion} gives some concluding remarks.

%%%%%%%%%%%%%%%%%%%%%%%%%%%%%%%%%%%%%%%%%%%%%%%%%%%%%%%%%%%%%%%%%%%%%%%%%%%%%%%%%%%%%%%%%%%%%%%%%%%%%%%%%%%%%%%%%%%%%%%%%%%%%%%%%%%%%%%%%%%%%%%%%%%%%%%%%%%%
\section{Skew Polynomials}
\label{s:skew}
\subsection{Notation}
Throughout this paper, we fix a finite field $\Fq$ and consider a finite field extension $\Fqm$ over $\Fq$. Let $\text{Aut}(\Fqm)$ be the automorphism group of $\Fqm$. We let $\sigma_s \in \text{Aut}(\Fqm)$ be such that $\sigma_s(a) = a^{q^s}$ for all $a \in \Fqm$. Since the maximal subfield fixed by $\sigma_s$ is $\Fq$ if and only if $\text{gcd}(s,m) = 1$, we will henceforth assume $\text{gcd}(s,m) = 1$ whenever we consider $\sigma_s \in \text{Aut}(\Fqm)$. Further, we denote the nonzero elements of $\Fqm$ by $\Fqm^*$ and we let $\mathbb{N} = \{0, 1, 2 \ldots \}$. 

For ease of presentation, for $i \in \mathbb{N}$, define ${\lsem i \rsem}_s = \frac{q^{is}-1}{q^s-1}$ and $[i]_s = q^{is}$. We can verify that ${\lsem i \rsem}_s$ and $[i]_s$ satisfy the following properties.
\begin{proposition}
\label{prop_q_properties}
For any $i,j \in \mathbb{N}$ and any $a \in \F_{q^m}$,
\begin{enumerate}
\item[(1)] $a^{[0]_s} = a$;
\item[(2)]$a^{[i]_s} = a^{[j]_s}$ if $i \equiv j \mod m$;
\item[(3)] $[i]_s[j]_s = [i+j]_s$;
\item[(4)]${\lsem i \rsem}_s + [i]_s =  {\lsem i + 1 \rsem}_s $;
\item[(5)] ${\lsem i \rsem}_s+[i]_s{\lsem j \rsem}_s = {\lsem i + j \rsem}_s.$
\end{enumerate}
\end{proposition}
When $s=1$ and there is no ambiguity, we will use the notation $\sigma, {\lsem i \rsem}, [i],$ suppressing the subscript $s$. 
%%%%%%%%%%%%%%%%%
\iffalse
\begin{proof}
\begin{enumerate}
\item $a^{q^m} = a$ is the field equation. 
\item Assume $i > j$ and write $i = j + km$ for some $k$. Then
\begin{align*}
 a^{[i]} = a^{q^{j+km}} =(a^{q^j})^{q^{km}} =a^{[j]},
\end{align*}
since $a^{q^{km}} = a$. 
\item $q^iq^j = q^{i+j}$.
\item \begin{align*}
{\lsem i+1 \rsem} -{\lsem i \rsem} = \frac{q^{i+1} - q^i}{q-1} = q^i = [i]
\end{align*}
\item \begin{align*}
{\lsem i \rsem}+[i]{\lsem j \rsem} = \frac{q^i-1}{q-1}+\frac{q^{i+j} -q^i}{q-1} 
= \frac{q^{i+j}-1}{q-1} = {\lsem i+j \rsem}. 
\end{align*}
\end{enumerate}
\end{proof}
\fi
%%%%%%%%%%%%%%%%

%%%%%%%%%%%%%%%%%%%%%%%%%%%%%%%
\subsection{Definition and Basic Properties}
\begin{definition}
The \emph{skew polynomial ring} over $\Fqm$ with automorphism $\sigma_s$, denoted $\Fqm[x;\sigma_s]$, is the ring which consists of polynomials $\sum_i c_ix^i$, $c_i \in \Fqm$, with the usual addition of polynomials and a multiplication that follows the commuting rule $xa = \sigma_s(a)x$. 
\end{definition}

\begin{remark}
Skew polynomial rings can be more generally defined over division rings \cite{Ore}. Our definition here is general in the case of finite fields. 
\end{remark}

\begin{example}
\label{ex_1}
Consider $\F_4[x;\sigma]$ with $\F_4 = \{0, 1,\alpha, \alpha^2 = 1 + \alpha\}$ and $\sigma(a) = a^2$. Then
\begin{align*}
(x+1)(\alpha x+ 1) &= x(\alpha x + 1) + (\alpha x + 1) \\
&= \sigma(\alpha)x^2 + x + \alpha x+ 1 \\
&= \alpha^2x^2 + \alpha^2x + 1. 
\end{align*}
\end{example}

Clearly, since $xa \neq ax$ in general, $\Fqm[x;\sigma_s]$ is generally a noncommutative ring. As the next example shows, it is \emph{not} a unique factorization domain. 
\begin{example}
Consider $\F_4[x;\sigma]$ as in Example \ref{ex_1}. Then 
\begin{align*}
x^4 + x^2 + 1 &= (x^2+x+1)(x^2 + x + 1) \\
&= (x^2 + \alpha^2)(x^2 + \alpha), 
\end{align*}
are two possible irreducible factorizations. 
\end{example}

However, $\Fqm[x;\sigma_s]$ is a \emph{right Euclidean domain} \cite{Jacobson}. This means that, for $f,g \in \Fqm[x;\sigma_s]$, there are unique $p, r \in \Fqm[x;\sigma_s]$ such that
\begin{align}
\label{eq:euclid}
f(x) = p(x)g(x) + r(x),
\end{align}
with either $r = 0$ or $\deg(r) < \deg(g)$. 
\begin{remark}
Since $\Fqm[x;\sigma_s]$ is a noncommutative ring, the order of $p(x)g(x)$ in Equation (\ref{eq:euclid}) is important. The name \emph{right} Euclidean domain refers to $g$ appearing to the \emph{right} of $p$. In fact, $\Fqm[x;\sigma_s]$ is also a \emph{left} Euclidean domain. 
\end{remark}
\begin{remark}
When $s=1$, as a ring, $\Fqm[x;\sigma]$ is isomorphic to the ring of polynomials over $\Fqm$ that are \emph{linearized} over $\Fq$ \cite{Lidl}. However, $\Fqm[x;\sigma]$ has a different evaluation map. 
\end{remark}

Since we have a well-defined division algorithm on $\Fqm[x;\sigma_s]$, the standard notion of greatest common divisor (gcd) and least common multiple (lcm) also have the corresponding generalizations.
\begin{definition}
For nonzero $f_1, f_2 \in \Fqm[x;\sigma_s]$, the \emph{greatest right common divisor (grcd)} of $f_1$ and $f_2$, denoted $\grcd(f_1,f_2)$, is the unique monic polynomial $g \in \Fqm[x;\sigma_s]$ of highest degree such that there exist $u_1, u_2 \in \Fqm[x;\sigma_s]$ with $f_1 = u_1g$ and $f_2 = u_2g$. 
\end{definition}
\begin{definition}
For nonzero $f_1, f_2 \in \Fqm[x;\sigma_s]$, the \emph{least left common multiple (llcm)} of $f_1$ and $f_2$, denoted $\llcm(f_1,f_2)$, is the unique monic polynomial $h \in \Fqm[x;\sigma_s]$ of lowest degree such that there exist $u_1, u_2 \in \Fqm[x;\sigma_s]$ with $h = u_1f_1$ and $h = u_2f_2$. 
\end{definition}
Using the division algorithm, we can easily verify the following.
\begin{proposition}
\label{prop:llcm_grcd}
For all $f_1, f_2 \in \Fqm[x;\sigma_s]$,
\begin{align*}
\deg(\llcm(f_1,f_2)) = \deg(f_1) + \deg(f_2) - \deg(\grcd(f_1,f_2)).
\end{align*}
\end{proposition}

The next proposition shows that for distinct $\sigma_{r_1}, \sigma_{r_2} \in \text{Aut}(\Fqm)$, the skew polynomial rings $\Fqm[x,\sigma_{r_1}]$ and $\Fqm[x,\sigma_{r_2}]$ are not isomorphic. 
\begin{proposition}
Let $\Fqm$ be a finite field and let $\sigma_{r_1}, \sigma_{r_2} \in \text{Aut}(\Fqm)$. Then the skew polynomial rings $\Fqm[x,\sigma_{r_1}]$ and $\Fqm[x,\sigma_{r_2}]$ are isomorphic as rings if and only if $\sigma_{r_1} = \sigma_{r_2}$. 
\end{proposition}
\begin{proof}
Suppose $\Psi : \Fqm[x,\sigma_{r_1}] \rightarrow \Fqm[x,\sigma_{r_2}]$ is a ring isomorphism. Clearly, $\Psi$ restricted to $\Fqm$ is an automorphism of the field, and $\Psi(x) = x$. Thus we have, on the one hand, for any $a\in \Fqm$,
\begin{align*}
\Psi(xa) = \Psi(\sigma_{r_1}(a)x) = \Psi(\sigma_{r_1}(a))x,
\end{align*}
and on the other,
\begin{align*}
\Psi(xa) = \Psi(x)\Psi(a) = x\Psi(a) = \sigma_{r_2}(\Psi(a))x. 
\end{align*}
Thus we need
\begin{align}
\label{eq:iso}
\Psi(\sigma_{r_1}(a)) = \sigma_{r_2}(\Psi(a)) \quad  \text{for all }  a \in \Fqm.  
\end{align}
Since $\Psi$ is an automorphism when restricted to $\Fqm$ and commutes with $\sigma_{r_1}$ and $\sigma_{r_2}$, (\ref{eq:iso}) holds if and only if $\sigma_{r_1} = \sigma_{r_2}$. 
\end{proof}

%%%%%%%%%%%%%%%%%%%%%%%%%%%%%%%
\subsection{$\sigma_s$-Conjugacy Classes}
To properly define the evaluation of skew polynomials, we need the concept of $\sigma_s$-conjugacy. We first consider the following map.
\begin{definition}
For $\sigma_s \in \text{Aut}(\Fqm)$, the \emph{$\sigma_s$-warping map} $\vp_{\sigma_s}$, is the map
\begin{align*}
    \vps: \Fqm^* &\longrightarrow \Fqm^*\\
    a &\longmapsto \sigma_s(a)a^{-1}. 
\end{align*}
\end{definition}
When $s=1$, we write $\vp$ for $\vp_\sigma$. 
\begin{proposition}
\label{prop:warp}
For $a,b \in \Fqm^*$, $\vps(a) = \vps(b)$ if and only if $a = bc$ for some $c \in \Fq^*$, i.e., if and only if $a$ and $b$ are in the same multiplicative coset of $\Fq^*$ in $\Fqm^*$.  
\end{proposition}
\begin{proof}
Observe that the map $\vps$ is multiplicative; and for $c \in \Fq^*$, $\vps(c) = 1$. Thus, $\vps(a) = \vps(bc)$. Conversely, if $\vps(a) = \vps(b)$, then $\vps(\frac{a}{b}) = 1$, showing $\frac{a}{b} \in \Fq^*$. 
\end{proof}
\begin{definition}
For any two elements $a \in \Fqm, c \in \Fqm^*$, define the \emph{$\sigma_s$-conjugation} of $a$ by $c$ as follows: 
\begin{align*}
a^c \stackrel{\triangle}{=} a \vps(c). 
\end{align*}
\end{definition}
\begin{definition}
We call two elements $a, b \in \Fqm$ \emph{$\sigma_s$-conjugates} if there exists an element $c \in \Fqm^*$ such that $a^c = b$.
\end{definition}
  
It is easy to verify that $\sigma_s$-conjugacy is an equivalence relation. We call the set $C_{\sigma_s}(a) = \{a^c \mid c \in \Fqm^*\}$ the \emph{$\sigma_s$-conjugacy class} of $a$. When $s=1$, we write $C(a)$ for $C_{\sigma_s}(a)$. 

\begin{corollary}
\label{lem:class_size}
For any $a \in \Fqm^*$, $|C_{\sigma_s}(a)| = {\lsem m \rsem}$. 
\end{corollary}
\begin{proof}
It follows from Proposition \ref{prop:warp} that there are exactly ${\lsem m \rsem}$ different values of $\vps(c)$ for $c \in \Fqm^*$. 
\end{proof}
\begin{proposition}
For any $a \in \Fqm$, we have that $C_{\sigma_s}(a) = C(a)$. 
\end{proposition}
\begin{proof}
Every element in $C_{\sigma_s}(a)$ has the form $a\vps(c)$ for some $c \in \Fqm^*$. Then,
\begin{align*}
a\vps(c) = ac^{q^s-1} = a(c^{\lsem s \rsem})^{q-1},
\end{align*}
which is in $C(a)$. Since by Corollary \ref{lem:class_size}, $C_{\sigma_s}(a)$ and $C(a)$ have the same size, $C_{\sigma_s}(a) = C(a)$. 
\end{proof}

\begin{example}
\label{ex_conj}
Consider $\F_{16}$, with a primitive element $\gamma$, and $\sigma(a) = a^4$. Then, $C(0) = \{0\}$ is a singleton set, and
\begin{align*}
C(1) 
%\{1^c \mid c \in \F_{16}^* \} \\
 = \{\vp(c) \mid c \in \F_{16}^* \} 
%& = \{(\gamma^i)^3 \mid 0\leq i \leq 14\} \\
 = \{1, \gamma^3, \gamma^6, \gamma^9, \gamma^{12} \}.
\end{align*}
Note that $C(1)$ is a subgroup of $\F_{16}^*$, while the other nontrivial classes are cosets of $C(1)$:
\begin{align*}
C(\gamma)  &= \{\gamma, \gamma^4, \gamma^7, \gamma^{10}, \gamma^{13} \}, \\
C(\gamma^2)  &= \{\gamma^2, \gamma^5, \gamma^8, \gamma^{11}, \gamma^{14} \}.
\end{align*}
\end{example}

In the previous example, we can use $1, \gamma, \gamma^2$ as class representatives. In general, there are $m-1$ nontrivial (excluding $C(0)$) $\sigma_s$-conjugacy classes for $\Fqm$ with $\sigma(a) =a^q$. Thus, we can use $\gamma^\ell$ with $0\leq \ell < m-1$ as the class representatives. 

%%%%%%%%%%%%%%%%%%%%%%%%%%%%%%%
\subsection{Skew Polynomial Evaluation}
To simplify the discussion of skew polynomials, we will often associate a skew polynomial in $\Fqm[x;\sigma_s]$ with two polynomials in $\Fqm[x]$ as follows.  
\begin{definition}
Let $f_s = \sum_i c_ix^i \in \Fqm[x;\sigma_s]$. Define $f^R_s, f^L_s \in \Fqm[x]$ as
\begin{align*}
f^R_s&= \sum_i c_ix^{\lsem i \rsem_s}, \\
f^L_s &= \sum_i c_ix^{[i]_s};
\end{align*}
we call $f^R_s$ and $f^L_s$ the \emph{regular associate} and \emph{linearized associate} of $f_s$, respectively. Moreover, we call any polynomial of the form $\sum_i c_ix^{[i]_s}$ an $s$-linearized polynomial. 
\end{definition}

When defining an evaluation map for a skew polynomial ring, it is important to take into account the action of $\sigma_s$. The traditional ``plug in'' map that simply replaces the variable $x$ by a value $a \in \Fqm$ does not work. A suitable evaluation map, using the fact that $\Fqm[x;\sigma_s]$ is a right Euclidean domain, was defined by Lam and Leroy \cite{Lam2}.
\begin{definition}
For $f \in \Fqm[x;\sigma_s], a \in \Fqm$, by right division, compute $f(x) = p(x)(x-a) + r$, with $r \in \Fqm$, and define the \emph{evaluation of $f$ at the point $a$} to be $f(a) = r$. 
\end{definition}

As the next theorem shows, we can compute this evaluation without using the division algorithm.
\begin{theorem}[Lam and Leroy]
\label{eval}
For $f_s = \sum_i c_ix^i \in \F_{q^m}[x;\sigma_s]$ and $a \in \F_{q^m}$, $f_s(a) = \sum_i c_i a^{{\lsem i \rsem}_s} = f^R_s(a)$. 
\end{theorem}
Thus, the evaluation of a skew polynomial  is equal to the evaluation of its regular associate. 
\begin{corollary}
\label{cor:root}
Zeros of $f_s \in \Fqm[x;\sigma_s]$ are in one-to-one correspondence with zeros of $f^R_s\in \Fqm[x]$. 
\end{corollary}

Unlike the evaluation map for ordinary polynomial rings,  this evaluation map is not a ring homomorphism. In particular, $fg(a) \neq f(a)g(a)$ in general. In order to evaluate a product, we need the previously-defined concept of $\sigma_s$-conjugacy class. 
\begin{theorem}[Lam and Leroy]
\label{thm:prod}
Let $f, g \in \Fqm[x;\sigma_s]$, and $a \in \Fqm$. If $g(a) = 0$, then $fg(a) = 0$, otherwise $fg(a) = f(a^{g(a)})g(a)$. 
\end{theorem}

\begin{example}
Consider $\F_4[x;\sigma]$ as before, with  $\sigma(a) = a^2$. Let $f = x^4 + x^2 + 1$, $g = x^2+x+1$ and $h =  x^2 + x + 1$, so that $f = gh$. By Theorem \ref{eval}, 
\begin{align*}
f(\alpha) = \alpha^{\lsem 4 \rsem} + \alpha^{\lsem 2 \rsem} + 1 = 1. 
\end{align*}
By Theorem \ref{thm:prod},
\begin{align*}
gh(\alpha) = g(\alpha^{h(\alpha)})h(\alpha) =\alpha^2 \alpha = 1.
\end{align*}
\end{example}

As the next theorem shows, the evaluation of skew polynomials is intimately related to the evaluation of linearized polynomials. 
\begin{theorem}
\label{t:lin_corr}
Let $f_s =\sum_{i=0}^nc_ix^i \in \Fqm[x;\sigma_s]$ and $f^L_s = \sum_{i=0}^nc_ix^{[i]_s} \in \Fqm[x]$ be the corresponding linearized associate. Then for any $a \in \Fqm$, 
\begin{align*}
af(\vp_{\sigma_s}(a)) = f^L_s(a). 
\end{align*}
\end{theorem}
\begin{proof}
\begin{align*}
a f(\vp_{\sigma_s}(a))&=a\left(\sum_{i=1}^nc_i(a^{q^s-1})^{\lsem i \rsem_s} \right)\\
    &= a\left(\sum_{i=0}^nc_i(a^{q^s-1})^{\frac{(q^s)^i-1}{q^s-1}}\right) \\
    &=\sum_{i=0}^nc_ia^{[i]_s} =f^L_s(a). 
\end{align*}
\end{proof}
When $s =1$, the linearized polynomial $f^L = \sum_{i=0}^nc_ix^{[i]} \in \Fqm[x]$ has at most $q^n$ roots, since, as a regular polynomial, it has degree at most $q^n$. The next theorem shows that the $s$-linearized polynomial $f^L_s = \sum_{i=0}^nc_ix^{[i]_s} \in \Fqm[x]$ has the same bound on the number of roots, even though it has a much higher degree when viewed as a regular polynomial. 
\begin{theorem} 
\label{thm:s_linear}
An $s$-linearized polynomial of degree $[n]_s$ in $\Fqm[x]$ has at most $q^n$ roots. 
\end{theorem}
\begin{proof}
We proceed by induction on $n$. For $n =0$, the polynomial $g_0 = a_0x$ with $a_0 \neq 0$ clearly has only one root at $x=0$.  
For $n \geq 1$, suppose $g_n$ is an $s$-linearized polynomial of degree $[n]_s$ and $\alpha \neq 0$ is a root of $g_n$. Since $g_n$ is linearized, for any $c \in \Fq$, $c\alpha$ is also a root of $g_n$. Thus, $g_n$ is divisible by the $s$-linearized polynomial $h = x^{q^s} - \alpha^{q^s-1}x$. Using the symbolic product of linearized polynomials \cite{Lidl}, we can express $g_n$ as $g_n = g_{n-1}(h(x))$, where $g_{n-1}$ is an $s$-linearized polynomial of degree $[n-1]_s$. By the induction hypothesis, $g_{n-1}$ has at most $q^{n-1}$ roots. Now for each root $\beta$ of $g_{n-1}$, since $\gcd(s,m)=1$, $h(x) = \beta$ has at most $q$ solutions. Thus, $g_n$ has at most $q^{n-1}q = q^n$ roots.
\end{proof}

%%%%%%%%%%%%%%%%%%%%%%%%%%%%%%%%%%%%%%%%%%%%%%%%%%%%%%%%%%%%%%%%%%%%%%%%%%%%%%%%%%%%%%%%%%%%%%%%%%%%%%%%%%%%%%%%%%%%%%%%%%%%%%%%%%%%%%%%%%%%%%%%%%%%%%%%%%%%
\section{Structure of $\sigma$-Conjugacy Classes}
\label{s:indep}

\subsection{Minimal Polynomials}
For any polynomial $f$, either in $\Fqm[x;\sigma_s]$ or in $\Fqm[x]$, let 
\begin{align*}
Z(f) = \{a \in \Fqm \mid f(a) =0 \}.
\end{align*}
That is, $Z(f)$ is the set of zeros of $f$. 

If $f \in \Fqm[x]$ is nonzero and $\deg(f) = n$, we know that $|Z(f)| \leq n$. However, as the next example shows, a skew polynomial can have more zeros than its degree.
\begin{example}
\label{ex:p_deg}
Let $f = x^2 + 1 \in \F_4[x;\sigma]$. Then, $Z(f) = \{1, \alpha, \alpha^2\}$, since, for $a \in \F_4^*$, 
\begin{align*}
f(a) &= a^{{\lsem 2 \rsem}} + 1 = a^3 + 1 = 0. 
\end{align*}
\end{example}

\begin{definition}
Let $\Omega \subseteq \Fqm$ and let $f_\Omega \in \Fqm[x;\sigma_s]$ be the monic polynomial of least degree such that $f_\Omega(a) = 0$ for all $a \in \Omega$. We call $f_\Omega$ the \emph{minimal polynomial} of $\Omega$. The empty set has $f_\emptyset = 1$. 
\end{definition}

\begin{proposition}
\label{prop:min_inc}
Let $\Omega \subseteq \Fqm$ and let $f_\Omega \in \Fqm[x;\sigma_s]$ be its minimal polynomial. Then for any $\beta \notin Z(f_\Omega)$, we have $f_{\Omega \cup \{\beta\}} = (x-\beta^{f_\Omega(\beta)})f_{\Omega}.$
\end{proposition}
\begin{proof}
By Theorem \ref{thm:prod}, we know that $(x-\beta^{f_\Omega(\beta)})f_{\Omega}$ vanishes on $\Omega \cup \{\beta\}$. To check minimality, we note that $\deg((x-\beta^{f_\Omega(\beta)})f_{\Omega}) = \deg(f_\Omega) + 1$ and no polynomial of $\deg(f_\Omega)$ can vanish on $\Omega \cup \{\beta\}$.  
\end{proof}

\begin{corollary}
\label{cor:min_lin}
Let $\Omega \subseteq \Fqm$. Then, $f_\Omega =(x-a_1)(x-a_2)\hdots(x-a_n)$ where each $a_i$ is conjugate to some element of $\Omega$. 
\end{corollary}
\begin{proof}
For any $\alpha \in \Omega$, $f_{\{\alpha\}} = x - \alpha$. The statement follows by iteratively applying Proposition \ref{prop:min_inc}. 
\end{proof}

Proposition \ref{prop:min_inc} and Corollary \ref{cor:min_lin} imply that the zeros of $f_\Omega$ are well-behaved in the following sense.
\begin{theorem}(Lam)
\label{thm:con_root}
Every root of $f_\Omega$ is a $\sigma$-conjugate to an element in $\Omega$.
\end{theorem}

We also state the following useful theorem.
\begin{theorem} (Lam and Leroy)
\label{thm:min_decomp}
Let $\Omega \subseteq \F$. If $f_\Omega = pg$, with $p,g \in \F[x;\sigma]$, then $g = f_{Z(g)}$, i.e., $g$ is a minimal polynomial. 
\end{theorem}

Lastly, we prove the following important decomposition theorem for minimal polynomials. 
\begin{theorem} [Decomposition Theorem]
\label{thm:llcm_grcd}
Let $\Omega_1, \Omega_2 \subseteq \F$, with corresponding minimal polynomials $f_{\Omega_1}$ and $f_{\Omega_2}$ such that $\Omega_1 = Z(f_{\Omega_1})$ and $\Omega_2 = Z(f_{\Omega_2})$. Then, the following holds
\begin{enumerate}
\item[(1)] $f_{\Omega_1 \cup \Omega_2} = \llcm(f_{\Omega_1}, f_{\Omega_2})$;
\item[(2)] $f_{\Omega_1 \cap \Omega_2} = \grcd(f_{\Omega_1}, f_{\Omega_2})$;
\item[(3)] $\deg(f_{\Omega_1 \cup \Omega_2}) = \deg(f_{\Omega_1}) + \deg(f_{\Omega_2}) -  \deg(f_{\Omega_1 \cap \Omega_2})$.
\end{enumerate}
\end{theorem}
\begin{proof}
\begin{enumerate}
\item[(1)] Since every $\alpha \in \Omega_1$ is a zero of $f_{\Omega_1 \cup \Omega_2}$, we have $f_{\Omega_1 \cup \Omega_2} = p_1 f_{\Omega_1}$ for some $p_1 \in \F[x;\sigma,\delta]$. Similarly, every $\beta \in \Omega_2$ is a zero of $f_{\Omega_1 \cup \Omega_2}$, so  we have $f_{\Omega_1 \cup \Omega_2} = p_2 f_{\Omega_2}$ for some $p_2 \in \Fqm[x;\sigma,\delta]$. Since $\llcm(f_1,f_2)$ is the monic polynomial of lowest degree with this property, we must have $f_{\Omega_1 \cup \Omega_2} = \llcm(f_{\Omega_1}, f_{\Omega_2})$.
\item[(2)] Every $\alpha \in \Omega_1 \cap \Omega_2$ is a zero of both $f_{\Omega_1}$ and $f_{\Omega_2}$. Thus, we can write $f_{\Omega_1} = p_1f_{\Omega_1 \cap \Omega_2}$ and $f_{\Omega_2} = p_2f_{\Omega_1 \cap \Omega_2}$ for some $p_1, p_2 \in \Fqm[x;\sigma,\delta]$. Thus, by the definition of grcd, we have $f_{\Omega_1 \cap \Omega_2} \mid \grcd(f_{\Omega_1},f_{\Omega_2})$, where $|$ denotes right divisibility. Now, clearly every $\beta \in Z(\grcd(f_{\Omega_1},f_{\Omega_2}))$ is a zero of both $f_{\Omega_1}$ and $f_{\Omega_2}$; since $\Omega_1 = Z(f_{\Omega_1})$ and $\Omega_2 = Z(f_{\Omega_2})$, we have that $\beta$ is a zero of $f_{\Omega_1 \cap \Omega_2}$. By Theorem \ref{thm:min_decomp}, $\grcd(f_{\Omega_1},f_{\Omega_2})$ is the minimal polynomial of $Z(\grcd(f_{\Omega_1},f_{\Omega_2}))$. Thus $f_{\Omega_1 \cap \Omega_2} = \grcd(f_{\Omega_1},f_{\Omega_2})$.
\item[(3)] Follows from $(1), (2)$ and Proposition \ref{prop:llcm_grcd}.
\end{enumerate}
\end{proof}

%%%%%%%%%%%%%%%%%%%%%%%%%%%%%%%%%%%%%%%%
\subsection{$P$-independent Sets}
Extending Example \ref{ex:p_deg}, we see that if $\Omega = \{1, \alpha \}$, then $f_\Omega = x^2 + 1$. However, $Z(f_\Omega) = \{1, \alpha, \alpha^2 \}$. This shows that, in a skew polynomial ring, it is possible that $|Z(f_\Omega)| > |\Omega|$. This motivates the following definition by Lam \cite{Lam1}. 

\begin{definition}
An element $\alpha \in \Fqm$ is \emph{$P$-dependent} on a set $\Omega$ if $f_{\Omega} = f_{\Omega \cup \{\alpha\}}$ and \emph{$P$-independent} of $\Omega$ otherwise. A set of elements $\Omega=\{\alpha_1,\dots,\alpha_n\}\subseteq \Fqm$ is $P$-independent if for any $i \in \{1,\dots,n \}$ the element $\alpha_i$ is $P$-independent of the set $\Omega\setminus\{\alpha_i\}$.
\end{definition} 

\begin{definition}
The \emph{$P$-closure} of the set $\Omega$ is 
\begin{align*}
 \overline{\Omega}=\{\alpha \in \Fqm \mid f_{\Omega}(\alpha)=0\}.
\end{align*}
Any maximal $P$-independent subset of $\overline{\Omega}$ is called a \emph{$P$-basis} for $\overline{\Omega}$. 
\end{definition}

An important theorem relating $P$-independent sets to $\sigma$-conjugacy classes is the following by Lam \cite{Lam1}. 
\begin{theorem} (Lam)
\label{thm:conj_indep}
Let $\Omega_1, \Omega_2 \subset \Fqm$ such that $\Omega_1$ and $\Omega_2$ are $P$-independent, and subsets of two distinct conjugacy classes. Then $\Omega = \Omega_1 \sqcup \Omega_2$ is also $P$-independent, where $\sqcup$ denotes a disjoint union. 
\end{theorem}

\begin{example}
\label{ex:F16}
Consider $\F_{16}$ with primitive element $\gamma$ and $\sigma(a) = a^4$. 
\begin{itemize}
\item The set $\{1, \gamma^3\}$ is $P$-independent. In fact, two element set is $P$-independent.
\item The set $\{1, \gamma^3, \gamma^6\}$ is not $P$-independent. In fact, $\overline{\{1, \gamma^3\}} = C(1)$.  
\item The set $\{1, \gamma^3, \gamma, \gamma^4\}$ is $P$-independent, as it is the disjoint union of $\{1, \gamma^3\} \in C(1)$ and $\{\gamma, \gamma^4 \} \in C(\gamma)$. 
\end{itemize}
\end{example}

Lam \cite{Lam1} also showed that the $P$-independence of a set can be determined by examining the degree of its minimal polynomial. 
\begin{theorem} [Lam]
\label{thm:P_size}
Let $\Omega\subseteq \Fqm$. Then $\Omega$ is $P$-independent if and only if $\deg(f_{\Omega})=|\Omega|$. 
\end{theorem}
In the following, we will require the following two useful corollaries. 
\begin{corollary}
\label{con_degree_2}
Let $\Omega =\{\alpha_1,\alpha_2, \ldots, \alpha_n\}$ be an arbitrary
set of $n$ points in $\Fqm$. Then $\deg(f_\Omega) \leq n$. 
\end{corollary}

\begin{corollary}
\label{subset_deg}
Let $\Omega = \{\alpha_1, \alpha_2,\ldots, \alpha_n\} \subset \F_{q^m}$ be
such that  $\deg(f_\Omega) = n$. Then for any $S \subset \Omega$, we have $\deg(f_S) = |S|$. 
\end{corollary}
\begin{proof}
Consider $\Omega_i = \Omega \setminus \{\alpha_i\}$ for each $1 \leq i \leq n$. By Corollary \ref{con_degree_2}, we know that $\deg(f_{\Omega_i}) < n$. Suppose that $\deg(f_{\Omega_i}) < n-1$, then the polynomial $g(x) = (x - \alpha_i^{f_{\Omega_i}(\alpha_i)})f_{\Omega_i}(x)$ vanishes on all of $\Omega$. However, $\deg(g) < \deg(f_\Omega) = n$, contradicting the minimality of $f_\Omega$. Thus, $\deg(f_{\Omega_i}) = n-1 = |\Omega_i|$ for every $i$. The result follows by recursively applying this argument for smaller subsets of $\Omega$. 
\end{proof}

%%%%%%%%%%%%%%%%%%%%%%%%%%%%%%%%%%%%%%%%
\subsection{Structure Theorem} 
When restricted to a single $\sigma$-conjugacy class, the $P$-independence structure of a set is related to linear independence. We now examine this connection.

\begin{lemma} [Independence Lemma]
\label{lem:indep}
Let $\Omega = \{\alpha_1, \ldots, \alpha_n\} \subseteq C_{\sigma_s}(\gamma^\ell) \subset \Fqm$, for $0 \leq \ell < m -1$, and $a_1,\dots,a_n\in \Fqm$ be such that $\alpha_i=\gamma^\ell \vps(a_i)$ for $i = 1, \ldots, n$. Then, $\Omega$ is $P$-independent if and only if $a_1, \ldots, a_n$ are linearly independent over $\F_q$. 
\end{lemma}
\begin{proof}
Without loss of generality, we assume $\ell=0$. Let $\Omega=\{\alpha_1,\dots,\alpha_n\}\subset C_{\sigma_s}(1)$ and $a_1,\dots,a_n\in \Fqm$ such that $\alpha_i= \vps(a_i)$ for all $i=1,\dots,n$. Let $f_\Omega  \in \Fqm[x;\sigma_s]$ be the minimal polynomial of $\Omega$ and $f_\Omega^L \in \Fqm[x]$ be the corresponding $s$-linearized associate. 
Let $\lambda_1,\dots,\lambda_n\in \F_q$ and $a=\sum_{i=0}^n\lambda_ia_i$ such that $a \neq 0$. By Theorem \ref{t:lin_corr},
\begin{align*}
a f_\Omega(\vp(a))&=f_\Omega^L(a) = \sum_{i=0}^n c_i a^{[i]_s} = \sum_{i=0}^n c_i \left (\sum_{j=0}^n \lambda_j a_j \right)^{[i]_s} \\
    &= \sum_{j=0}^n  \lambda_j \sum_{i=0}^n c_i (a_j)^{[i]_s} = \sum_{j=0}^n \lambda_j f_\Omega^L(a_j) = 0,
\end{align*}
where the last equality follows from the fact that for each $j$, $a_j f_\Omega(\alpha_j) = f_\Omega^L(a_j)$ and $f_\Omega(\alpha_j) = 0$. This shows that for every $a \in Z(f_\Omega^L)$, $\vp(a) \in \overline{\Omega}$.

If $a_1, \ldots, a_n$ are linearly independent, then by Theorem \ref{thm:s_linear} $\deg(f_\Omega^L) \geq [n]_s$. Thus, $\deg(f_\Omega) \geq n$. By Corollary \ref{con_degree_2}, $\deg(f_\Omega)$ is at most $n$. Therefore, $\deg(f_\Omega)= n$ and $\Omega$ is $P$-independent by Theorem \ref{thm:P_size}. 

Conversely, assume $\Omega$ is $P$-independent. Without loss of generality, suppose $a_n$ is linearly dependent on $\{a_1, \ldots, a_{n-1} \}$. The above calculation shows that $\alpha_n$ is a root of $f_{\Omega \setminus \{a_n\}}$. This contradicts the $P$-independence assumption. 
\end{proof}

\begin{corollary}
\label{cor:indep}
Let $\Omega \subseteq C_{\sigma_s}(1) \subset \Fqm$, for $0 \leq \ell < q -1$, be a $P$-independent set.  Then, $\alpha$ is a root of $f_\Omega$ if and only if $\alpha = \vps(a)$, where $a$ is a root of $f_\Omega^L$.
\end{corollary}
\begin{proof}
The proof of the Independence Lemma shows that if $a$ is a root of $f_\Omega^L$, then $\vps(a)$ is a root of $f_\Omega$. The converse follows from Theorem \ref{thm:con_root} and Theorem \ref{t:lin_corr}. 
\end{proof}

\begin{corollary}
\label{cor:closed_size}
Let $\Omega = \{\alpha_1, \ldots, \alpha_n\} \subset C_{\sigma_s}(\gamma^\ell) \subset \Fqm$ be a $P$-independent set, for some $0 \leq \ell < m -1$. Then $|\overline{\Omega}| = \lsem n \rsem$.
\end{corollary}
\begin{proof}
Using the proof of the Independence Lemma, we see that the restriction of the warping map, $\vps :Z(f_\Omega^L) \setminus \{0\} \rightarrow Z(f_\Omega)$ is a $(q-1)$ to $1$ map. The independence assumption implies that $|Z(f_\Omega^L)| = q^n$. Corollary \ref{cor:closed_size} shows that the restriction of the warping map is onto.  Thus $|Z(f_\Omega)| = \frac{q^n-1}{q-1} = {\lsem n \rsem}$.
\end{proof}
\begin{remark}
In case $s=1$, $f_\Omega$ has degree $n$ and its regular associate $f^R_\Omega$ has degree ${\lsem n \rsem}$. This shows that $f^R_\Omega$ splits in $\Fqm$. However, when $s \neq 1$, the corresponding $f^R_\Omega$ has degree ${\lsem n \rsem}_s$, but only has ${\lsem n \rsem}$ roots over $\Fqm$. 
\end{remark}

\begin{theorem} [Structure Theorem]
\label{p_indep}
Let $\Omega = \{\alpha_1, \ldots, \alpha_n\} \subseteq C_{\sigma_s}(\gamma^\ell) \subset \Fqm$, for $0 \leq \ell < m -1$, and $a_1,\dots,a_n\in \Fqm$ be such that $\alpha_i=\gamma^\ell \vps(a_i)$ for $i = 1, \ldots, n$. Then
\begin{align}
\label{eq:corr} 
 \overline{\Omega}=\{\gamma^\ell \vps(a) \mid a\in \langle a_1,\dots,a_n\rangle\}\subseteq C_{\sigma_s}(\gamma^\ell)\
\end{align}
where $\langle a_1,\dots,a_n\rangle$ denotes the $\F_q$ subspace of $\Fqm$ generated by $\{a_1, \ldots, a_n\}$. 
\end{theorem}

\begin{proof}
In light of the Independence Lemma and Corollary \ref{cor:indep}, it suffices to show that, without loss of generality, if $\alpha_1, \ldots, \alpha_k$ is a $P$-basis for $\Omega$, then $\langle a_1, \ldots, a_k  \rangle = \langle a_1, \ldots, a_n \rangle$. Now for any  $a \in \langle a_1, \ldots, a_n \rangle$, the calculation in the proof of Independence Lemma shows that $a \in Z(f_\Omega^L)$. Since $\alpha_1, \ldots, \alpha_k$ are $P$-independent, we know that  $\deg(f_\Omega) = k$ and thus $\deg(f_\Omega^L) = [k]_s$. Since $ a_1, \ldots, a_k $ is linearly independent, we see that $Z(f_\Omega^L)= \langle a_1, \ldots, a_k  \rangle$. Thus, $\langle a_1, \ldots, a_k  \rangle = \langle a_1, \ldots, a_n \rangle$. 
\end{proof}

\begin{remark}
The Structure Theorem can also be derived from the work of Lam and Leroy \cite{Lam2}. Here we presented a direct approach and drew the important connection to linearized polynomials. 
\end{remark}

%%%%%%%%%%%%%%%%%%%%%%%%%%%%%%%%%%%%%%%%%%%%%%%%%%%%%%%%%%%%%%%%%%%%%%%%%%%%%%%%%%%%%%%%%%%%%%%%%%%%%%%%%%%%%%%%%%%%%%%%%%%%%%%%%%%%%%%%%%%%%%%%%%%%%%%%%%%%
\section{Matroidal Structure } 
\label{s:matroid}
\subsection{Matroid Basics}
In the following, we will only give the basics of matroid theory and follow the notation given in \cite{Oxley}. All the important results in this subsection can be found in \cite{Oxley} and are only restated here for completeness. 

\begin{definition}
A \emph{matroid} $M$ is an ordered pair $(E, \cI)$, where $E$ is a finite set and $\cI$ is a set of subsets of $E$ satisfying the following three conditions:
\begin{enumerate}
\item[\bf(I1)] $\emptyset \in \cI$;
\item[\bf(I2)] If $I \in \cI$ and $I' \subseteq I$, then $I' \in \cI$; 
\item[\bf(I3)] If $I_1, I_2 \in \cI$ and $|I_1| < |I_2|$, then there is an element $e \in I_2 - I_1$ such that $I_1 \cup \{e\} \in \cI$. 
\end{enumerate}
\end{definition}
If $M = (E, \cI)$ is a matroid, then $M$ is called a matroid on $E$. The members of $\cI$ are called the \emph{independent sets} of $M$ and $E$ is called the \emph{ground set} of $M$. 

A simple class of matroids is defined as follows. 
\begin{definition}
Let $E = \{1, \ldots n\}$ and let $0 \leq m \leq n$.  For any subset $X \in E$, declare $x \in \cI$ if and only if $|X| \leq m$. Then $M = (E, \cI)$ is called the $(n,m)$-\emph{uniform matroid} and is denoted by $U_{n,m}$. 
\end{definition}

An important class of matroids comes from linear algebra.
\begin{definition}
Let $A$ be an $m \times n$ matrix over a field $\F$. Let $E = \{1, \ldots, n \}$. For any $X \subseteq E$, $X \in \cI$ if the columns indexed by $X$ are linearly independent over $\F$. The pair $(E, \cI)$  forms a matroid called the \emph{vector matroid} of $A$. 
\end{definition}

\begin{example}
\label{ex:vm}
Let 
\begin{align*}
A =
 \begin{pmatrix}
  1 & 0 & 0 & 1 \\
  0 & 1 & 0 & 1 \\
  0 & 0 & 1 & 0
 \end{pmatrix}
\end{align*}
be a $3 \times 4$ matrix over $\F_2$. Then $E = \{1,2,3,4\}$ and $\cI = \{ \emptyset, \{1\}, \{2\}, \{3\}, \{4\}, \{1,2\},$ $ \{1,3\}, \{1,4\},$ $ \{2,3\}, \{2,4\},\{3,4\}, \{1,2,3\}, \{1,3,4\}, \{2,3,4\}\}.$
\end{example}

Two matroids $(E_1, \cI_1)$ and $(E_2, \cI_2)$ are \emph{isomorphic} if there exists a bijection $f: E_1 \rightarrow E_2$ such that $I \in \cI_1$ if and only if $f(I) \in \cI_2$. 

\begin{definition}
A matroid $M$ is \emph{representable over a field $\F$} (\emph{$\F$-representable}) if it is isomorphic to the vector matroid of some matrix over $\F$. A matroid is \emph{representable} if it is representable over some field. 
\end{definition}

\begin{definition}
Let $M$ be a matroid. A maximal independent set in $M$ is a \emph{basis} of $M$. 
\end{definition}
It is easy to see that all bases of a matroid $M$ have the same size.

\begin{example}
In Example \ref{ex:vm}, the sets $\{1,2,3\}, \{1,3,4\},$ $\{2,3,4\}$ are all bases of $(E, \cI)$. 
\end{example}

Let $M$ be the matroid $(E, \cI)$ and let $X \subseteq E$. Let $\cI| X = \{I \subset X: I \in \cI\}$. Then the pair $(X, \cI|X)$ is a matroid. We call this matroid the restriction of $M$ to $X$, and denote it by $M|X$. 
\begin{definition}
The \emph{rank} $r(X)$ of $X$ is the size of a basis of $M|X$. 
\end{definition}
It can be verified the rank function $r$ satisfies the following:
\begin{enumerate}
\item[\bf(R1)]  If $X \subseteq E$, then $0 \leq r(X) \leq |X|$; 
\item[\bf(R2)]  If $X \subseteq Y \subseteq E$, then $r(X) \leq r(Y)$;
\item[\bf(R3)]  If $X, Y \subseteq E$, then 
\begin{align*}
r(X \cup Y) + r(X \cap Y) \leq r(X) + r(Y). 
\end{align*}
\end{enumerate}
Conversely, as the following theorem shows, conditions \textbf{(R1)-(R3)} characterize the rank function of a matroid.
\begin{theorem} 
Let $E$ be a set and $r$ be a function that maps $2^E$ into the set of non-negative integers and satisfies \textbf{(R1)-(R3)}. Let $\cI$ be the collection of subsets $X$ of $E$ for which $r(X) = |X|$. Then $(E, \cI)$ is matroid having rank function $r$. 
\end{theorem}

\begin{definition}
Let $M = (E, \cI)$ be a matroid, for any $X \subseteq E$, define the \emph{closure} of $X$, denoted $\text{cl}(X)$, as
\begin{align*}
\text{cl}(X) = \{x \in E \mid r(X \cup x) = r(X) \}. 
\end{align*}
If $X = cl(X)$, then $X$ is called a \emph{flat}. 
\end{definition}

Let $\cF(M)$ be the set of all flats of a matroid $M = (E, \cI)$. Furthermore, for any $X \subseteq E$, let
\begin{align*}
\cF(X) = \{U \subseteq X \mid U = \text{cl}(U) \},
\end{align*}
i.e., $\cF(X)$ denotes the set of all flats contained in $X$. 

\begin{example}
In Example \ref{ex:vm}, $\{1,3\}, \{2, 3\}$ are flats. However, $\{1, 2\}$ is not a flat as $\text{cl}(\{1,2\}) = \{1,2,4\}$. We have that $\cF(\{1,2,3\}) = \{\emptyset, \{1\}, \{2\}, \{3\}, \{1,2\}, \{2,3\} \}$. 
\end{example}

\subsection{The $\Fqm[x;\sigma]$-matroid}
For the rest of the paper, we will restrict to the case $s=1$ and consider the ring $\Fqm[x;\sigma]$. We shall see, in light of the Structure Theorem, that we do not lose generality with this restriction. 

\begin{theorem}
Let $\Fqm[x;\sigma]$ be a skew polynomial ring. Then the pair $M=(\Fqm, \mathcal{I})$, where 
\begin{align*}
\mathcal{I}=\{\Omega\subseteq \Fqm \mid |\Omega|=\deg(f_{\Omega})\}
\end{align*}
is the set of all $P$-independent sets of $\Fqm$, is a matroid.
\end{theorem}

\begin{proof}
Nonzero constant polynomials have no roots, thus $\emptyset\in \mathcal{I}$. Suppose $I\in \mathcal{I}$ and let $I'\subset I$. From Corollary \ref{subset_deg}, $I'$ is $P$-independent set. 

Now let $I_1,I_2\in \mathcal{I}$ with $|I_1|<|I_2|$.We need to prove that there exists an element $e\in I_2\setminus I_1$ such that $I_1\cup\{e\}$ is still a $P$-independent set. Suppose to the contrary that for all $e\in I_2\setminus I_1$ it holds that $I_1\cup\{e\}\not\in \mathcal{I}$. It follows that $I_2$ is $P$-dependent on $I_1$. This contradicts the fact that  $|I_1|<|I_2|$ and $I_2 \in \mathcal{I}$. 
\end{proof}

We can easily verify the following correspondences between notions in matroid theory and notions defined in terms of $P$-independence. 
\begin{lemma}\label{l:rank}
Let $M = (\Fqm, \cI)$ be the matroid constructed from $\Fqm[x;\sigma]$ and let $X \subset M$. Then
\begin{itemize}
\item $X$ is an independent set in $M$ if and only if $X$ is a $P$-independent subset of $\Fqm$;  
\item $\text{cl}(X)$ is equal to the $P$-closure of $X$; 
\item $\deg(f_X)$ is a rank function on $M$. 
\end{itemize}
\end{lemma}

\begin{theorem}
\label{thm:m_rep}
$M=(\Fqm, \mathcal{I})$ is an $\F_q$-representable matroid. 
\end{theorem}
\begin{proof}
Fix a basis of $\Fqm$ over $\F_q$ and represent each element of $\Fqm$ as a column vector over $\F_q$. Consider a class $C(\gamma^\ell) = \{\alpha_1, \ldots, \alpha_{\lsem m \rsem} \}$. For any $\alpha_i \in C(\gamma^\ell)$, we can find $a_i$ such that $\alpha_i = \gamma^\ell a_i^{q-1}$. Consider the $m \times {\lsem m \rsem}$ matrix over $\F_q$
\begin{align*}
A =  \begin{pmatrix}
  a_1 & a_2 & \ldots & a_{\lsem m \rsem} \\
 \end{pmatrix}.
\end{align*}
By Theorem \ref{p_indep}, any subset of columns of $A$ are linearly independent over $\F_q$ if and only if the corresponding $\alpha_i$'s are $P$-independent. Thus, the column linear independence structure of $A$ exactly represent the $P$-independence structure of $C(\gamma^\ell)$. 

Since union of $P$-independent sets from distinct classes remain $P$-independent, we can consider the following construction.
Let $\cA$ be a $(m(q-1) + 1) \times ({\lsem m \rsem}(q-1) + 1)$ matrix given by:
\begin{align*}
\cA= \begin{pmatrix}
 A_1 & 0 & \ldots & 0& 0 \\
 0 & A_2 & \ldots & 0& 0 \\
 \vdots & \vdots& \ddots & \vdots & \vdots \\
 0 & 0 & \ldots & A_{q-1}& 0 \\
 0 & 0 & \ldots & 0 & 1 \\
\end{pmatrix},
\end{align*}
where each $A_\ell = A$ for $0 \leq \ell \leq q-1$, and the last column is a column of ${\lsem m \rsem}(q-1)$ zeros followed by a $1$. Clearly if we associate the columns in $A_\ell$ with the class $C(\gamma^\ell)$ and the last column with the class $C(0) = \{0\}$, then the linear independence structure of the columns of $\cA$ will correspond to the $P$-independence structure of $\Fqm$. Thus $M=(\Fqm, \mathcal{I})$ is an $\F_q$-representable matroid. 
\end{proof}

\begin{example}
Consider $\F_{16}$ with primitive element $\gamma$ and $\sigma(a) = a^4$. Let $M=(\F_{16}, \mathcal{I})$. Let $\{1, \gamma \}$ be a basis of $\F_{16}$ over $\F_4$, where $\F_4^* = \{1, \gamma^5, \gamma^{10}\}$. Then the vector $(1, \gamma,  \gamma^2, \gamma^3, \gamma^4 )$ expands into a $2 \times 5$ $A$ matrix over $\F_4$ as
\begin{align*}
A &= \begin{pmatrix}
 1 & 0 & \gamma^5 & \gamma^5 &1  \\
 0 & 1 &  1 & \gamma^{10} & 1 \\
\end{pmatrix}.
\end{align*}
The matrix
\begin{align*}
\cA= \begin{pmatrix}
 1 & 0 & \gamma^5 & \gamma^5 & 1 & 0 & 0 & 0 & 0 & 0 & 0 & 0 & 0 & 0 & 0 & 0\\
 0 & 1 &  1 & \gamma^{10} & 1& 0& 0 & 0 & 0 & 0 & 0 & 0 & 0 & 0 & 0 & 0 \\
 0 & 0 & 0 & 0 & 0 & 1 & 0 & \gamma^5 & \gamma^5 & 1& 0 & 0 & 0 & 0 & 0 & 0\\
 0 & 0 & 0 & 0 & 0 & 0 & 1 &  1 & \gamma^{10} & 1 & 0 & 0 & 0 & 0 & 0 & 0\\
 0 & 0 & 0 & 0 & 0 &  0 & 0 & 0 & 0 & 0 & 1 & 0 & \gamma^5 & \gamma^5 & 1 & 0 \\
 0 & 0 & 0 & 0 & 0 &  0 & 0 & 0 & 0 & 0 &  0 & 1 &  1 & \gamma^{10} & 1 & 0 \\
 0 & 0 & 0 & 0 & 0 &  0 & 0 & 0 & 0 & 0 & 0 & 0 & 0 & 0 & 0 & 1 \\
 \end{pmatrix}
\end{align*}
is an $\F_4$-representation of $M$. 
\end{example}

\begin{remark}
The representation we gave in the proof of Theorem \ref{thm:m_rep} is the most ``efficient'' representation of $M = (\Fqm, \mathcal{I})$ over $\F_q$ in the sense that the associated $\cA$ matrix has the smallest dimension over $\F_q$. Indeed, the largest independent set in $M$ has size $m(q-1)+1$, which corresponds to the number of \emph{rows} of $\cA$. 
\end{remark}

%%%%%%%%%%%%%%%%%%%%%%%%%%%%%%%%%%%%%%%%
\subsection{$\cF(\Fqm)$ Metric Space}
Let $\cF(\Fqm)$ denote the set of all flats in the $\Fqm[x,\sigma]$-matroid. We now show that $\cF(\Fqm)$ is a metric space. 
\begin{theorem}
Define the map
\begin{align*}
 d_\cF: \cF(\Fqm) \times  \cF(\Fqm) &\longrightarrow \mathbb{N}\\
  (X, Y)&\longmapsto r(X\cup Y)-r(X\cap Y).
\end{align*}
Then, $d_\cF$ is a metric on $\cF(\Fqm)$. 
\end{theorem}
\begin{proof}
Since symmetry and non-negative definiteness are obvious, it suffices to show that $\dcF$ satisfies the triangle equality. Let $X,Y,Z \in \cF(\Fqm)$. We want to show that 
\begin{align*}
\dcF(X,Y) - \dcF(X,Z) - \dcF(Y,Z) \leq 0. 
\end{align*}
By Theorem \ref{thm:llcm_grcd}, we know that 
\begin{align*}
\dcF(X,Y) = \deg(f_X) + \deg(f_Y) - 2\deg(f_{X\cap Y}). 
\end{align*}
Thus, 
\begin{align*}
& \dcF(X,Y) - \dcF(X,Z) - \dcF(Y,Z) =\\
& \quad = 2\deg(f_{X\cap Z}) + 2\deg(f_{Y \cap Z}) - 2\deg(f_Z) - 2\deg(f_{X \cap Y}) \\
&\quad = 2\deg(f_{(X \cap Z)\cup(Y \cap Z)}) + 2\deg(f_{X \cap Y \cap Z})  - 2\deg(f_Z) - 2\deg(f_{X \cap Y}) \\
&\quad = 2(\deg(f_{(X \cap Z)\cup(Y \cap Z)}) - \deg(f_Z))   + 2(\deg(f_{X \cap Y \cap Z})  - \deg(f_{X \cap Y})) 
\leq  0,
\end{align*}
since both $\deg(f_{(X \cup Z)\cap(Y \cup Z)}) - \deg(f_Z)\leq 0$ and $\deg(f_{X \cap Y \cap Z})  - \deg(f_{X \cap Y}) \leq 0$. 
\end{proof}
Thus $\cF(\Fqm)$ together with the map $d_\cF$ is a metric space. We shall denote it as $(\cF(\Fqm),d_\cF)$.

%%%%%%%%%%%%%%%%%%%%%%%%%%%%%%%%%%%%%%%%
\subsection{The $C(1)$-submatroid and Projective Geometry}
From the matroid representation in Theorem \ref{thm:m_rep}, it is easy to see that any single conjugacy class of $\Fqm$ is itself a representable matroid. Since all nontrivial classes have the same structure, we shall examine $C(1)$. Denote $\cF(C(1))$ as the set of all flats of the $C(1)$-submatroid. Clearly the restriction of $\dcF$ to $\cF(C(1))$ makes $(\cF(C(1)),d_\cF)$ a metric space. We now show the correspondence between $(\cF(C(1)),d_\cF)$ and the projective geometry of vector space $\Fqm$ over $\F_q$. 

Viewing $\Fqm$ as a vector space over $\F_q$, let $\cP(\Fqm)$ denote the set of all nontrivial subspaces of $\Fqm$. Then, as shown in \cite{Kschischang}, the \emph{subspace metric}, $d_S$, defined for all $V, W \in \cP(\Fqm)$ as
\begin{align*}
d_S(V,W) = \dim (V + W) - \dim (V \cap W),
\end{align*}
is a metric on $\cP(\Fqm)$. 

Let $(\cP(\Fqm), d_S)$ be the metric space $\cP(\Fqm)$ with the subspace metric. We arrive at the following correspondence theorem. 
\begin{definition}
Define the \emph{extended warping map}, $\Phi$, between the metric spaces $(\mathcal{P}(\Fqm),d_S)$ and $(\cF(C(1)),d_\cF)$, via 
\begin{align*}
\Phi:\cP(\Fqm)&\longrightarrow \cF(C(1))\\
V&\longmapsto \{\vp(a)\mid a\in V \setminus \{0\}\}. 
\end{align*}
\end{definition}

\begin{theorem}
\label{thm:isometry}
$\Phi$ is a bijective isometry. 
\end{theorem}
\begin{proof}
We first show the map is injective. Let $V_1,V_2\in\cP(\Fqm)$ be such that $V_1\neq V_2$. Let $a\in V_1\setminus V_2$. By Proposition \ref{prop:warp}, it follows that $\vp(a)\notin V_2$. Therefore $\vp(a) \in V_1 \setminus V_2$, so $\Phi$ is injective.

For surjectivity, let $\{\alpha_1, \ldots, \alpha_n\}$ be a $P$-basis for a flat in $\cF(C(1))$. By the Independence Lemma, there exist $a_1, \ldots a_n \in \Fqm$ such that $\vp(a_i)= \alpha_i$ for all $i$, and $\{a_1, \ldots, a_n\}$ is linearly independent over $\F_q$. Thus $\langle a_1, \ldots, a_n \rangle \in \cP(\Fqm)$. 

To show isometry, note that for $V, W \in \cP(\Fqm)$, $\dim(V + W) = \dim(V) +\dim(W) - \dim(V \cap W)$. Clearly, $\dim(V) = r(\Phi(V))$. Thus, in light of Theorem \ref{thm:llcm_grcd}, it suffices to show $\dim(V \cap W) = r(\Phi(V) \cap \Phi(W))$. Towards this end, let $a_1, \ldots, a_j$ be a basis for $V \cap W$. Clearly $\vp(a_1), \ldots, \vp(a_j) \in \Phi(V) \cap \Phi(W)$. By the Independence Lemma, $\vp(a_1), \ldots \vp(a_j)$ are $P$-independent. Thus $\dim(V \cap W) \leq r(\Phi(V) \cap \Phi(W))$. Conversely, if $\alpha_1, \ldots, \alpha_k$ is a $P$-basis for $\Phi(V) \cap \Phi(W)$, then there exist linearly independent $a_1, \ldots, a_k \in V \cap W$. This shows $\dim(V \cap W) \geq r(\Phi(V) \cap \Phi(W))$. Thus, $\dim(V \cap W) = r(\Phi(V) \cap \Phi(W))$. 
\end{proof}

%%%%%%%%%%%%%%%%%%%%%%%%%%%%%%%%%%%%%%%%%%%%%%%%%%%%%%%%%%%%%%%%%%%%%%%%%%%%%%%%%%%%%%%%%%%%%%%%%%%%%%%%%%%%%%%%%%%%%%%%%%%%%%%%
%%%%%%%%%%%%%%%%

\section{Application to Matroidal Network Coding}
\label{s:comm}

Network coding, introduced in the seminal paper \cite{ACLY00},
is based on the simple idea that, in a packet network, intermediate
nodes may forward \emph{functions} of the packets that they receive,
rather than simply routing them.  Using network coding, rather
than just routing, greater
transmission rates can often be achieved.
In \emph{linear} network coding, packets are interpreted as vectors
over a finite field, and intermediate nodes
forward linear combinations of the vectors that they receive.
Sink nodes receive such linear combinations, and are able
to recover the original message provided that they can solve
the corresponding linear system.
In \emph{random} linear network coding (RLNC), the linear combinations
are chosen at random, with solvability of the linear system
assured with high probability when the underlying field is
sufficiently large \cite{HO06}.

As a means of introducing error-control coding in RLNC,
recognizing that random linear combinations of vectors
are subspace-preserving,
K\"{o}tter and Kschischang
\cite{Kschischang} introduced the concept of transmitting
information over a network encoded in subspaces.
In this framework, 
the packet alphabet is the set of all vectors of a
vector space, and the message alphabet is the set of
all subspaces of that space.
The source node encodes a message in a subspace and transmits
a basis of that space.
Each intermediate
node then forwards a random linear combination of
its incoming packets.
Each sink collects incoming packets and
reconstructs the subspace that was selected at the transmitter. 

Gadouleau and Goupil \cite{Gadouleau} generalized
the subspace framework to a matroidal one.
In this framework, the packet alphabet is the ground set of a matroid,
and the message alphabet is the set of all flats of that matroid.
The source node
encodes a message in a flat of the matroid and transmits a basis of that flat.
Each intermediate
node then forwards a random element of the flat
generated by its incoming packets.
Each sink collects incoming packets and
reconstructs the flat that was selected at the transmitter.
In our work, we will use the
$\Fqm[x;\sigma]$-matroid in this matroidal network coding framework. 

\subsection{Communication using $\Fqm[x;\sigma]$-matroid}

We first consider using only the $C(1)$-submatroid. The setup can be summarized as the following.

\begin{itemize}
\item The packet alphabet is $C(1)$ and the message alphabet
is $\mathcal{F}(C(1))$.
\item The source node encodes a message into a flat $\Omega$ of $C(1)$ and
  sends a basis of $\Omega$.
\item An intermediate node receives $\alpha_1,\dots,\alpha_h\in \Omega$
 and
  forwards a random root of the minimal polynomial $f_{\{\alpha_1,\dots,\alpha_h\}}\in \Fqm\xs$.
\item Each sink node collects sufficiently many packets to generate $\Omega$.
\end{itemize}

\begin{remark}
As a consequence of Theorem \ref{thm:isometry}, this $C(1)$-submatroid
communication model has the same message alphabet size
as the subspace communication
model and has the packet size of the projective network coding
model in \cite{Gadouleau}.
\end{remark}

We can extend the message alphabet size
in the $C(1)$-submatroid setup as follows.

\begin{itemize}
\item The message alphabet is
\[
\mathcal{F}(C(1)) \cup
\mathcal{F}(C(\gamma)) \cup \cdots \cup
\mathcal{F}(C(\gamma^{q-2}))
\]
and the packet alphabet is $\Fqm^*$.
\item The source node encodes a message into a flat
$\Omega_\ell \in \mathcal{F}(C(\gamma^\ell))$. 
\item An intermediate node receives $\alpha_1,\dots,\alpha_h\in \Omega_\ell$
and
forwards a random root of the minimal
 polynomial $f_{\{\alpha_1,\dots,\alpha_h\}}\in \Fqm\xs$.
\item Each sink node collects sufficiently many packets to
generate $\Omega_\ell$.
\end{itemize}
This setup increases the message alphabet size by a factor of $q-1$.

\begin{remark}
In both cases above, we could have included the $C(0)$-submatroid.
This amounts to sending the zero packet at the source, which each
intermediate node simply forwards. 
\end{remark} 

\subsection{Computational Complexity}
\label{s:complexity}
The computation at an intermediate node in $\Fqm\xs$ matroid network
coding is considerably more
complex than that of subspace transmission. In the latter case,
an intermediate node simply needs
to compute a random linear combination of the incoming packets.
In the $\Fqm[x;\sigma]$-matroidal scheme,
 an intermediate node must forward a random  root of
the minimal polynomial of its incoming packets. Following the Structure
Theorem, this can be accomplished as follows.

Let $\alpha_1,\dots,\alpha_h\in C(\gamma^\ell)$ be the incoming packets at an
intermediate node.  Note that all incoming packets are elements of the same
class;  the intermediate node must first determine this class (which we call
the \emph{Class Membership} problem).  Next, the intermediate node can find
$a_i\in \Fqm$ such that $a_i^{q-1}=\alpha_i\gamma^{-\ell} \in C(1)$ for
$i=1,\dots,h$ (which we call the \emph{Root Finding} problem).  Finally, the
intermediate node can compute a random nonzero
$\F_q$-linear combination $a\in \langle
a_1,\dots,a_h\rangle$, and then forward $\alpha=\gamma^\ell a^{q-1}\in
C(\gamma^\ell)$.  Since the complexity of the last two tasks is well-known, we
shall focus on the complexity of the first two. 

\subsubsection{Class Membership}

Without loss of
generality, we focus on the first received packet
$\alpha_1\in C(\gamma^\ell)$.  It holds that
$\alpha_1=\gamma^\ell a_1^{q-1}$ for some $a_1\in \F_{q^m}$. It is
possible to isolate the parameter $\ell$ by using the following
exponentiation:
\begin{align*}
\alpha_1^{\lsem m\rsem}&=\gamma^{\ell \lsem m\rsem}a_1^{(q-1)\lsem
  m\rsem}\\
&=\left(\gamma^{\lsem m\rsem}\right)^\ell\in \F_q^*.
\end{align*} 
The class membership problem can then be solved by means of an exponentiation
by $\lsem m\rsem$ and the use of a look-up table for a reasonably
small parameter $q$.

\subsubsection{Root Finding}

We propose two different approaches.
The first one is general and based on solving a multivariate linear
system of equations over $\F_q$,  while the second method is more efficient,
but only works in specific field extensions. 

\paragraph{Method 1}
For $\alpha\in C(1) \subset \Fqm$, we can compute a $(q-1)$-th root of $\alpha$
by solving the equation $x^{q-1} - \alpha= 0$. This is equivalent to finding a
nonzero root of the polynomial $x^q - \alpha x$. Since $x^q - \alpha x$ is a
linearized polynomial, this amounts to solving a linear system with $m$
equations over $\Fq$; using Gaussian elimination this can be
done using $O(m^3)$ operations over $\F_q$. 

\paragraph{Method 2} 
Let $\Fqm$ be an extension of $\F_q$ such that
$\gcd(\lsem m \rsem, q-1) = 1$. Given $\alpha = a^{q-1} \in \Fqm$,
find $t$ such that $(q-1)t = 1 \mod {\lsem m \rsem}$, and
compute $\alpha^t= a^{(q-1)t} = a$. 
Note that $q-1$ is invertible modulo ${\lsem m \rsem}$ if and only if
$\gcd(\lsem m \rsem, q-1) = 1$. Thus, our condition on the field extension size
is necessary. Furthermore, $t$ can be precomputed since the field extension is
fixed. Computing $\alpha^t$ takes $O(\log t)$ multiplications in
$\Fqm$. Assuming each multiplication is $O(m\log m)$ complexity in $\F_q$, the
overall algorithm takes $O((\log t) m\log m)$ complexity in $\F_q$. 

%%%%%%%%%%%%%%%%%%%%%%%%%%%%%%%%%%%%%%%%%%%%%%%%%%%%%%%%%%%%%%%%%%%%%%%%%%%%%%%%%%%%%%%%%%%%%%%%%%%%%%%%%%%%%%%%%%%%%%%%%%%%%%%%%%%%%%%%%%%%%%%%%%%%%%%%%%%%%%%%%%%%%%
\section{Conclusions}
\label{s:conclusion}

In this work, we highlighted the connection between the evaluation of skew
polynomials and that of linearized polynomials. Using linearized polynomials,
we gave a simple proof of a structure theorem relating $P$-independence for
skew polynomials and linear independence for vector spaces. This structure
theorem  allows us to construct the $\Fqm[x;\sigma]$-matroid. Using a
decomposition theorem for minimal polynomials, we showed that the
$\Fqm[x;\sigma]$-matroid is a metric space. Furthermore, the $C(1)$-submatroid
is bijectively isometric to projective geometry of $\Fqm$ equipped with the
subspace metric. Using this isometry, we showed that the
$\Fqm[x;\sigma]$-matroid can be used in a matroidal network coding framework.

\section*{Acknowledgments}

The work of Siyu Liu and Frank R. Kschischang
was supported by a Discovery Grant from the
Natural Sciences and Engineering Research Council (NSERC), Canada.
The work of Felice Manganiello was supported
by the Schweizerischer Nationalfonds (SNF), Switzerland.

\end{document}